# Applications for Microwave Kinetic Induction Detectors in Advanced Instrumentation

**Gerhard Ulbricht** [1,2,*], **Mario De Lucia** [1,2] and **Eoin Baldwin** [1,2]

1   Dublin Institute for Advanced Studies, School of Cosmic Physics, 31 Fitzwilliam Place, Dublin 2, D02XF86, Ireland; delucia@cp.dias.ie (M.D.L.); baldwin@cp.dias.ie (E.B.)
2   Trinity College Dublin, School of Physics, College Green, Dublin 2, Ireland
*   Correspondence: ulbrichtg@cp.dias.ie

**Abstract:** In recent years Microwave Kinetic Inductance Detectors (MKIDs) have emerged as one of the most promising novel low temperature detector technologies. Their unrivaled scalability makes them very attractive for many modern applications and scientific instruments. In this paper we intend to give an overview of how and where MKIDs are currently being used or are suggested to be used in the future. MKID based projects are ongoing or proposed for observational astronomy, particle physics, material science and THz imaging, and the goal of this review is to provide an easily usable and thorough list of possible starting points for more in-depth literature research on the many areas profiting from kinetic inductance detectors.

## 1. Introduction

Sensors to detect and characterize particles or photons are key components for many areas in modern science. Currently, the most successful particle or photon detectors are semiconductor-based CCD and CMOS arrays. They are used intensively for high energy particles and photons but are especially dominant for light around the ultraviolet (UV) and optical where Si-based arrays have achieved impressive results and immense array sizes. As their detection principle is to excite charge carriers across the semiconductor band gap, this band gap defines the detector's basic characteristics and is responsible for significant limitations. For example, the band gap in Si is around 1.1 eV, thus any photon to be detected with a Si-based CCD needs at least this 1.1 eV of energy (equivalent to a 1088 nm wavelength) to excite one electron, and even a photon with just a little below twice this energy still results in the same signal. Semiconductor-based CCDs therefore can only achieve energy resolution on a single pixel basis at energies that are significantly higher than their band gap. For this reason, small band gap semiconductors like e.g., HgCdTe, are used for lower photon energies in the infrared (IR) but practical limitations for their smallest achievable band gaps still restrict their energy resolution and low energy cut-off.

An intriguing option to build on and extend the tremendous success of semiconductor-based detectors is to instead use the much smaller band gaps found in superconductors. The well-known effect of superconductivity of a vanishing resistance below a critical temperature $T_c$ is caused by the pairing of charge carrying electrons into Cooper pairs which then condense into the lowest possible state. The superconductor band gap $\Delta$ separates this lowest state from higher up ones, and any Cooper pair that gets excited over this gap is broken into unpaired and dissipative electrons, forming a so-called quasi-particle (QP). Superconductor band gaps can be significantly smaller compared to typical semiconductor gaps and regarding detectors, superconductors with small $T_c$ are preferred—as a small $T_c$ also means a small band gap $\Delta$. Aluminum, for example, can have a $T_c$ of 1.2 K and a $\Delta$ of $1.8 \cdot 10^{-4}$ eV. As $\Delta$ is defined per electron, an energy of at least $2\Delta$ is required to break a Cooper pair, but this is still almost four orders of magnitude below the Si band gap. This allows superconducting detectors to achieve stronger signals, higher sensitivities for lower energies, and much better energy resolution. Superconductors come with further advantages like the lack of dissipation and high charge carrier mobility, but they also have one obvious disadvantage: they have to operate at very low temperatures to stay superconductive. This comes with the added benefit of vanishing thermal noise but can present significant engineering challenges, especially if many pixels in a large detector array are desired.

Cooling power at temperatures below 1 K is typically very limited, and every electrical connection required between the detector array and room temperature unavoidably contributes to the thermal load on the low temperature stage. If every single low temperature detector in an array requires two or more electrical connections this can severely restrict the possible number of pixels. (CCD arrays avoid millions of wires by applying a voltage across the array and shifting their pixel rows one by one downwards into the readout row, combined with reading that row pixel by pixel by again stepwise shifting the individual signals. This cannot be done with superconductors as due to their vanishing resistance voltages can be applied). It is therefore important for a low temperature detector to be multiplexable, allowing many individual detectors to be monitored with only a few electrical connections. Among the different superconducting photon and particle detectors developed so far, Microwave Kinetic Inductance Detectors (MKIDs, or sometimes just Kinetic Inductance Detectors, KIDs) [1] have the most straightforward and most efficient way of multiplexing (please see chapter 2), promising scalability up to mega-pixel low temperature detector arrays. Therefore, MKIDs allow to exploit the significant advantages of superconducting low temperature detectors in arrays with (currently) thousands to tens of thousands of pixels, allowing them to become one of the most promising novel detector technologies. They are being developed and adapted for more and more modern scientific applications in astronomy, particle physics, and many other areas.

The goal of this paper is not to give an exhaustive and complete overview of all MKID literature published so far as this would easily lead to an overwhelming, impracticable, and not very helpful review. Instead we intend to give an as complete as possible overview of how and where MKIDs are being used so far or are suggested to be used soon and to highlight several of the most important publications on these applications. The goal is to provide an easily usable, audited, and complete list of possible starting points for further literature research for anyone who wants to read deeper into the topic.

## 2. Microwave Kinetic Inductance Detectors: Basic Operation Principle

Every charge carrier (electrons, holes, or Cooper pairs) has a finite mass, and due to interactions with the surrounding lattice its effective mass often differs from the mass of a free electron. The term 'kinetic inductance' describes that (caused by this final charge carrier mass) it always takes time and energy to reverse the direction of motion of charge carriers. Therefore, when an AC voltage is applied, current will always trail behind voltage, like it does in a coil due to magnetic inductance. The size of this additional inductance in a given circuit depends on the charge carrier's effective mass and velocity, and thus on charge carrier density. It is low for normal metals with high electron densities and a short free mean path, but high for superconductors (the Cooper pair density is low in comparison), or at high AC frequencies.

In a superconductor, the interaction with the lattice leads to an attractive force between electrons, causing them to form pairs at very low temperatures. These Cooper pairs are no longer Fermions but Bosons, and thus will condense into the lowest state, allowing charge to be transported without resistance. The distance in energy to the next highest state is called the superconducting band gap $\Delta$. Breaking a Cooper pair is equivalent to lifting its two electrons to the next higher state and thus requires at least twice this band gap energy. The resulting two unpaired electrons are called a quasi-particle and will recombine into a Cooper pair after the material specific quasi-particle lifetime.

MKIDs use the effect of kinetic inductance in superconductors for photon or particle detection: For an MKID, a superconductor is lithographically patterned into an LC resonant circuit (see Figure 1a), a harmonic oscillator with a well-defined resonant frequency given by its capacitance C and total inductance $L_{total} = L_{geometric} + L_{kinetic}$. (Figure 1a shows a schematic for an LC-resonator with dedicated and separated inductor $L$ and capacitor $C$. For many applications MKIDs follow this design using an interdigitated capacitor and a long thin superconducting line as inductor. These MKIDs are often called Lumped Element KIDs, or LEKIDs to distinguish them from quarter-wave resonators. Quarter-wave MKIDs are preferred for some applications and use a superconducting line floating at one end and grounded at the other as resonator. They have their own advantages and disadvantages for detector array design and layout but follow the same basic operation principles.) If a photon with an energy of more than twice the superconducting band gap hits this resonator, some of the photon's energy will be transferred into heat and the rest will break Cooper pairs. This breaking of Cooper pairs by photon absorption reduces the LC-resonator's charge carrier density and thus increases the velocity of the remaining Cooper pairs that still have to carry the same current. This leads to an increase in kinetic inductance and thus a reduction of the LC-resonator's resonant frequency $f$. This change in resonant frequency can be measured very precisely and is the signal that is monitored with an MKID to detect photons or particles.

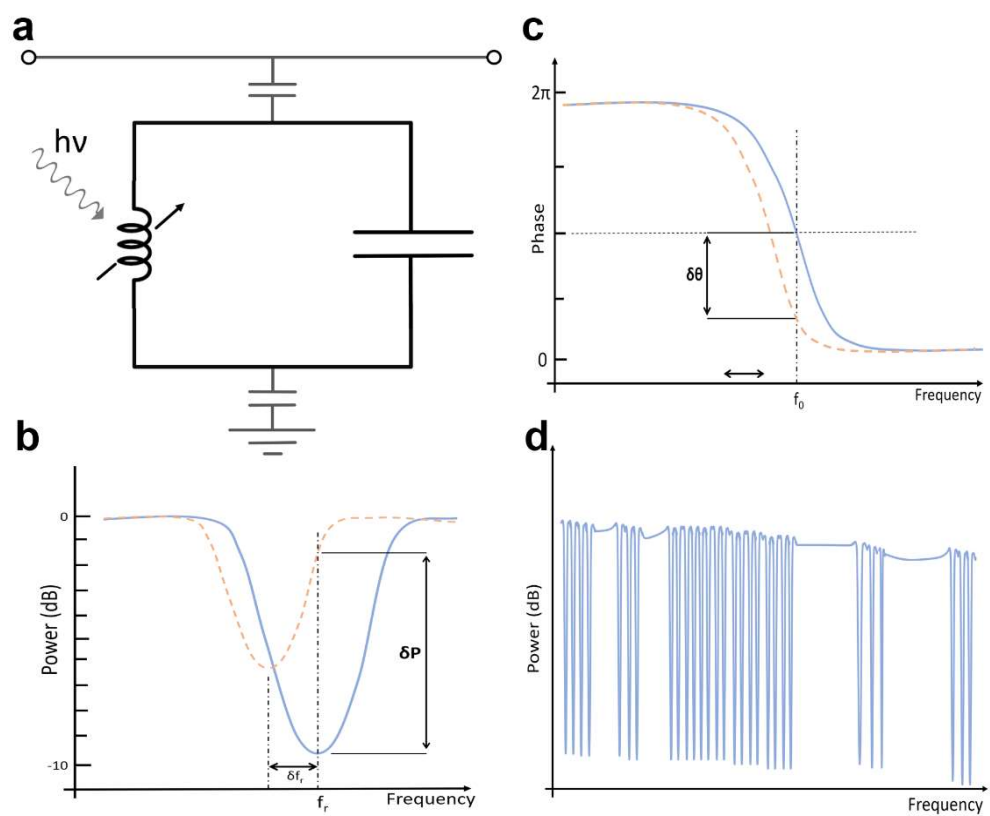

**Figure 1.** MKID operation principle. (a) LC resonant circuit; (b) amplitude of transmitted feedline signal before (blue) and after (red) photon absorption; (c) feedline signal phase response; (d) several resonators multiplexed with a single feedline. At resonance the *LC* circuit is effectively shorting the capacitively coupled microwave feedline to ground, resulting in a sharp dip in transmitted power.

MKID resonators are capacitively coupled to a microwave feedline (which is why they are called 'microwave' kinetic inductance detectors) and constantly driven at their resonant frequency. The change in $f_r$ caused by photon absorption therefore shifts both the amplitude (Figure 1b) as well as the phase response (Figure 1c) of the signal transmitted on the feedline to lower frequencies. As MKID *LC*-resonators are superconducting they have very little electrical losses and thus a very sharp resonance. Even a tiny change in resonant frequency produces a significant change in the phase signal (the sketch of Figure 1b for the amplitude signal is exaggerated for clarity, the change in amplitude is often much smaller). This phase shift can be quickly and very precisely measured, and every single photon that hits an MKID produces its own signal pulse. The pulse height is determined by the amount of broken Cooper pairs and thus the photon's energy. For sufficient photon energy this is used to not only count single photons but also to determine their individual energies. For longer wavelengths, single photons that can still break Cooper pairs no longer cause measurable signal pulses, but the photon flux can still be measured by monitoring the shift of the MKID's phase signal as the quasi-particle concentration will reach a flux dependent equilibrium between Cooper pair breaking and QP recombination.

As they are superconducting detectors MKIDs do have to operate at mK temperatures in most cases. As available cooling power at below 1 K is small, the heat influx from the surrounding room temperature environment has to be minimized. It is therefore important to be able to monitor and read out as many pixels as possible on a single electrical line between room temperature and the detectors themselves. This is the unique strength of MKIDs among low temperature detectors: they offer build-in frequency domain multiplexing. Every MKID pixel in a detector array has its individual, lithographically defined resonant frequency, usually spaced evenly over the frequency range the readout electronics can cover. As superconducting *LC*-resonators have very sharp resonances they do barely attenuate signals that are even just a few hundred kHz off their resonance. Many MKID resonators can thus be coupled to the same microwave feedline (see Figure 1d) without significant crosstalk. If one detector is hit by a photon and shifts in frequency the others will not react. This way it is possible to couple to and monitor thousands of MKIDs with only a few feedlines, requiring just two electrical connections to room temperature each. Up to 2000 MKID pixels per feedline have already been demonstrated [2,3] and MKID arrays with up to 20,440 individual detectors are being used at the time of writing [3]. The current limit of 2000 pixels per feedline is the result of the highest acceptable single detector resonant frequency (currently 8 GHz due to lack of very low noise amplifiers for higher frequencies), the fact that harmonic frequencies have to be avoided (resulting in a 4–8 GHz frequency window), and a minimum distance of 2 MHz between MKIDs in frequency space to optimize fabrication yield and time resolution.

The goal of this chapter was to only give a very short overview of the operating principle of Microwave Kinetic Inductance Detectors. A more thorough discussion would go beyond the scope of this paper but can be found for example in several excellent reviews [4–6].

## 3. MKID Applications in Observational Astronomy

As MKIDs have originally been developed for astronomical instrumentation [1,7] it is not surprising that their most established application so far is in observational astronomy, detecting and characterizing photons. We have divided this chapter in two parts for low energy photons (mid-IR and longer wavelengths) and higher ones (near-IR, optical and shorter wavelengths) for better clarity. But the working principle of MKIDs in these ranges is also slightly different. For higher energies, single photons break enough Cooper pairs to cause detectable single photon signal spikes, allowing single photon counting and energy resolution. For lower energies, starting around 1.5 µm it becomes increasingly difficult to identify single photon events. Instead, MKIDs are operated in a way (see below) that their signal represents the energy flux impacting the detector. This way, photon energy and photon flux become indistinguishable—less photons of higher energy will result in the same signal then more photons with less energy—on a single pixel basis, but the detector's wavelengths limit is increased significantly.

### 3.1. Observational Astronomy: Infrared to Microwaves

Especially for low energy, high wavelengths photons, MKIDs are a very capable detector. Their small superconductor band gap allows them to be sensitive down to very small photon energies, they are simple to fabricate compared to alternatives like semiconducting bolometers or transition edge sensors and their unrivaled multiplexability facilitates large detector arrays as required by many modern scientific instruments.

In this lower energy part of the electromagnetic spectrum single photons striking an MKID do not cause a detectable signal on their own. Instead, to generate a signal a sufficiently high photon flux is required to constantly break enough Cooper pairs. A steady flux results in a constant quasi-particle generation rate which will reach an equilibrium with QP recombination resulting in a constant QP density, depending on the number of photons per second and their energies. In this regime a single MKID pixel cannot distinguish between photon flux and wavelength and filters are often used in front of the detector to break this degeneracy. However, it is still possible to achieve energy resolution with MKIDs even for these low energies by using more than one detector in an on-chip spectrometer.

#### 3.1.1. MKID on-Chip Spectrometers

The DEep Spectroscopic HIgh-redshift MApper (DESHIMA) [8–10] uses this working principle. DESHIMA is a single pixel microwave spectrometer sensitive between 332 and 377 GHz (903–795 µm). It uses an on-chip microwave antenna on which incoming light is focused on in order to couple incoming photons into a superconducting feedline made from NbTiN. This feedline is coupled to 49 individual MKIDs, each with its own planar superconducting passband filter between MKID and feedline. This way only frequencies corresponding to the passband filter can travel from the feedline to the MKID. As a part of these MKIDs is fabricated from Al, which has a smaller gap $\Delta$ then NbTiN, photons that travel along the feedline without dissipation still break Cooper pairs in the Al and thus produce a signal in the MKID proportional to the photon flux within the filter's bandwidth. DESHIMA with its 49 frequency channels reaches a spectral resolution $f/\Delta f$ of 380. As the single pixel of the DESHIMA spectrometer only covers a few cm² on the chip these could be combined into a compact, multi-pixel spectral imager in the future [8]. DESHIMA has already been used for first demonstration observations with the 10 m diameter ASTE (Atacama Submillimeter Telescope Experiment) telescope in Pampa La Bora, Chile [8,9].

The design of an on-chip filter-bank spectrometer utilizing MKIDs is also pursued by SuperSpec [11–13]. SuperSpec is targeting the 195–310 GHz (1.54–0.97 mm) frequency range and is aiming for up to 500 channels to achieve a spectral resolution R of 700 [12]. Development is ongoing and prototypes with 50 channels and an R of 275 in the 255–278 GHz (1.18–1.07 mm) range have been demonstrated [11]. SuperSpec uses a Nb feedline and MKIDs made from sub-stoichiometric $TiN_x$ and is planned to be used soon with the Large Millimeter Telescope (LMT) Alfonso Serrano, Sierra Negra, Mexico. It is also proposed to be used as the detector element in X-Spec [14], a multi-object spectrometer for the proposed Cerro Chajnantor Atacama Telescope (CCAT), Chile.

Targeting a frequency range of 103–114.7 GHz (2.9–2.6 mm) is the CAMbridge Emission Line Surveyor (CAMELS) [15–17], another on-chip filter-bank spectrometer using a NbN feedline and MKIDs with a sensitive part made from β-phase Ta [17]. CAMELS intends to utilize 512 filter channels to reach a spectral resolution R of 3000 and is planned to operate at the 12 m Greenland Telescope. A further MKID instrument, WSPEC [18] intends to cover the 135–175 GHz (2.2–1.7 mm) and 190–250 GHz (1.6–1.2 mm) bands with R = 200. WSPEC is not using on-chip superconducting filter banks but a normal-metal, rectangular waveguide splitting into 54 branches each with a precisely machined filter structure and each terminated with an MKID made from Al.

Micro-Spec (μ-Spec) [19–22] is an MKIDs based project designed to measure spectra in the 0.5–1.0 mm (300–600 GHz) range using a different principle. Quite simplified, light from the telescope is focused onto an antenna structure that couples the photons into a Nb feedline. This feedline is split up into up to 256 arms [20], each with its own phase delay structure, realized by microstrip transmission lines of varying lengths. These 256 feedlines end in feed horns on one side of an on-chip, 2-dimensional parallel-plate waveguide region. They emit light into that region where it forms an interference pattern on the other side. Where the interference pattern is formed up to 345 receiving feed horns are positioned in an arc and couple the light into a further set of feedlines. Each is connected to a bank of order-sorting filters and its own Al or Mo$_2$N MKID [22,23] that are read out in order to measure the diffraction pattern and extract the spectrum from it. A spectral resolution between 512 [20] and 1200 [21] is expected. Micro-Spec is designed to be a highly miniaturized diffraction grating spectrometer that would fit on a 10 cm² Si chip. Due to this significant size and weight advantage it is proposed to replace huge and heavy, classic far-IR diffraction grating spectrometers in balloon based and space missions. The balloon born instrument EXCLAIM (the EXperiment for Cryogenic Large-Aperture Intensity Mapping) [23] plans to use six Micro-Spec spectrometers to do a low spatial resolution intensity mapping of the night sky in the 420–540 GHz (710–560 μm) range.

### 3.1.2. MKID Mid-IR to Microwave Imagers

To build an astronomical camera for the mid-IR to microwave range multiple MKID resonators have to be arranged as pixels in a detector array located in the focal plane of a telescope. Different methods are used to focus and couple the incoming light onto the MKID resonators themselves where it can break Cooper pairs. The incoming energy flux can then be measured via the QP density of each individual MKID pixel and an image can be reconstructed.

One of the first projects to use MKIDs for scientific instrumentation has been the photometric imaging camera MUSIC (MUltiwavelength Sub-millimeter Inductance Camera) and its precursor and test bed DemoCam, both intended for the Caltech Submillimeter Observatory (CSO) on Mauna Kea, Hawaii. DemoCam [24,25] could detect two bands at 240 and 350 GHz (1.2 and 0.9 mm) with 16 pixels. Each pixel had a slot antenna to couple incoming light to a microstrip-line, which then split, passed through band defining on-chip, superconducting bandpass filters and then coupled to two Al/Nb MKIDs per pixel. DemoCam was used for tests and to gather general know-how for MUSIC but was also used for first astronomical observations [25,26]. MUSIC [27–31] was intended to reach 576 individual pixels sensitive to 4 different colors at 150, 230, 290, and 350 GHz (2.0, 1.3, 1.0, and 0.86 mm). MUSIC used the same broadband, superconducting phased-array slot-dipole antennas coupled to Al/Nb MKIDs as DemoCam, just with more bandpass filters and 4 MKIDs per pixel for a total detector count of 2304. MUSIC was used for several scientific observation runs during its commissioning phase [29,32] but was decommissioned after the Caltech Submillimeter Observatory was closed. That also put the further development of MAKO [33] on hold, a 500 pixel MKID sub-mm camera prototype for the CSO to study scaling up to larger arrays using TiN based LEKIDs.

A very successful application of MKIDs as mm wavelength camera is the NIKA and NIKA2 collaboration. The New IRAM KIDs Array (NIKA) [34–37] is a 400 pixels cryogenic camera for astronomical observations with the 30 m IRAM (Institute for Millimetric Radio Astronomy) telescope on Pico Veleta, Spain. NIKA originally could observe in 2 frequency bands: 125–175 GHz (2.4–1.7 mm) with a 144 pixels array and 200–280 GHz (1.5–1.1 mm) with a larger 256 pixels array. The band selection is achieved with a cold dichroic and the Al MKIDs use fractal Hilbert-curve shaped inductors as receiving antenna for incoming mm wavelength light. NIKA has been operating successfully since 2013 and has been up-dated with its successor NIKA-2 in 2015. NIKA-2 [37–40] covers two similar bands centered around 150 GHz (2.0 mm) and 260 GHz (1.15 mm) and offers a 1040 pixels array for 150 GHz and two arrays with 1200 pixels each for both polarization directions of the 260 GHz band, for a total of 3440 MKIDs. NIKA-2 is openly accessible for users of the IRAM telescope since 2017 and together with NIKA has already produced significant scientific output like [40–43] to list just a few examples.

The Large Millimeter Telescope (LMT) Alfonso Serrano, Sierra Negra, Mexico, is the world's largest single aperture telescope for mm wavelengths, and two projects involving MKID cameras are ongoing at the LMT. MUSCAT, the Mexico UK Sub-mm Camera for AsTronomy [44,45] is a collaborative effort to install a 1600 pixels MKID camera at the LMT with a sensitivity band centered around 270 GHz (1.1 mm). MUSCAT uses aluminum MKID pixels that are insensitive to the incoming light's polarization and utilizes a meandering inductor optimized to absorb at the 1.1 mm wavelength. TolTEC [46,47] is the second MKID camera project for the LMT. It is intended as millimeter-wave imaging polarimeter with three bands around 1.1, 1.4, and 2.0 mm (280, 220, and 150 GHz), selected by dichroic filters. It is proposed to have 1900, 950, and 475 pixels in its 1.1, 1.4, and 2.0 mm band respectively, and with two MKIDs per pixel (one per polarization direction) TolTEC will have 6650 MKIDs in total. MKIDs have also been an integral part in A-MKID [48–50], the APEX MKID camera, a project to build a large format, two band (850 μm/353 GHz and 350 μm/855 GHz) camera for the 12 m Apex (Atacama Pathfinder Experiment) telescope, Llano de Chajnantor Observatory, Chile. Efforts are also ongoing to build a 110 pixels, 90–110 GHz (3.3–2.7 mm) MKID camera for the 45 m telescope of Nobeyama Radio Observatory near Minamimaki, Japan. And there have been proposals published for the multicolor sub-THz KID-array camera MUSICAM [51], targeting four 25 pixel-arrays for 2.0, 1.33, 1.04, and 0.86 mm (150, 225, 288, and 350 GHz) and intended for the proposed Eurasian Sub-Millimeter Telescopes (ESMT) and for TeSIA [52], an MKID camera with 1024 pixels covering the 850 GHz (350 μm) band for the proposed Date5 far-IR telescope on Dome A, Antarctica.

A long-time goal for the MKID based on-chip filter-bank spectrometers (described in chapter 3.1.1) is to be combined into larger arrays to allow high resolution imaging spectroscopy of astronomical sources in the mm and sub-mm wavelength ranges. With between 50 and 500 individual MKIDs per pixel these arrays would have a very large detector count. A different approach to achieve at least medium energy resolution in this frequency range is to use a Martin-Puplett Interferometer (MPI) like the one suggested for the CONCERTO [53] instrument. CONCERTO, the CarbON CII line in post-rEionisation and ReionisaTiOn epoch instrument is a proposed MKID imaging spectrometer for the 12 m Apex telescope. CONCERTO intends to utilize two 2152 pixels MKID arrays similar to the NIKA-2 design to detect both polarizations of incoming light in the focal plane of its MPI. A Martin-Puplett Interferometer (MPI) is a Fourier transform spectrometer which resembles a Michelson interferometer with an oscillating mirror at the end of one arm and can deliver a differential spectrum with respect to a given reference. With an oscillation frequency of 4 Hz, CONCERTO expects to achieve a resolution of 1 GHz across its 130–310 GHz (2.3–0.97 mm) band for a spectral resolving power R between 130 and 300 [53]. KISS (the KIDs Interferometer Spectrum Survey) [54,55] is a pathfinder instrument for CONCERTO and has recently been commissioned to the 2.25 m QUIJOTE telescope of Teide Observatory, Tenerife. KISS uses two 316 pixels arrays with MKIDs made from Ti/Al bilayers, observes in the 80–300 GHz (3.7–1.0 mm) band and has an MPI with a mirror path length of 10 cm and mirror oscillation frequency of 5 Hz. It reaches 1 GHz frequency resolution and is mainly dedicated (like CONCERTO) to observe galaxy clusters by studying secondary anisotropies of the Cosmic Microwave Background (CMB).

In the far-IR to microwave region of the electromagnetic spectrum Earth's atmosphere absorbs most of the incoming light. Many astronomical observations can therefore only be performed from space with telescopes on satellites. There are several projects to adapt MKIDs to be used as far-IR to microwave detectors for in-space applications. The SPACEKIDs collaboration [56–59] aimed to investigate the general suitability of MKIDs for far-IR to sub-mm space applications and demonstrated among others a 961 pixels MKID array centered around 850 GHz (350 μm) with competitive noise levels and cosmic ray dead times. MKIDs made from Al and NbTiN have been suggested [60,61] for the far-IR (34–210 μm) spectrometer SAFARI on the proposed 2.5 m SPICA (Space Infrared telescope for Cosmology and Astrophysics) infrared space telescope, and prototype arrays of 989 membrane-suspended MKIDs for the 1.4–2.8 THz (214–107 μm) band have been demonstrated [58]. MKIDs are also being considered for CORE [62], the Cosmic ORigins Explorer, a space telescope to observe the CMB polarization. CORE is proposed to cover 60–600 GHz (5–0.5 mm) in 19 bands with a 1.2 m telescope and 2100 pixels. For the Origins Space Telescope (OST) a concept for highly sensitive Al MKIDs that would be capable to count single mid-IR photons has been proposed [63,64]. OST is a mid-IR to far-IR telescope concept with a 6–9 m aperture and a targeted sensitivity band of 6–40 μm (50–7.5 THz). The JAXA (Japan Aerospace Exploration Agency) satellite for CMB observations LiteBIRD also initially considered MKIDs for its main instrument [65,66]. MKID development work is ongoing for GEP, the Galaxy Evolution Probe and its pathfinder balloon-born mission GEP-B [67]. GEP is a concept for an in-space observatory to study mid- and far-IR spectra of galaxies, designed with large arrays with more than 25,000 detectors. MKIDs for GEP should be able to cover 10–400 μm (30–0.75 THz), and especially in the 10–100 μm range few MKID designs have so far been demonstrated.

Compared to satellites, high altitude balloons offer an easier and less cost intensive alternative to place a telescope above most of Earth's atmosphere. BLAST-TNG [68–70], the next generation upgrade for the 'Balloon-borne Large Aperture Submillimeter Telescope' is using a 2.5 m telescope on a balloon to study the role magnetic fields play in star formation by observing polarized light from the interstellar medium in three bands centered around 250, 350, and 500 μm (1200, 857, and 600 GHz). For this BLAST-TNG uses MKIDs made from TiN in large arrays of 1836, 938, and 544 detectors for the 250, 350, and 500 μm bands respectively. A single BLAST-TNG pixel has two MKIDs, one for each polarization, with their inductors at 90 degrees to each other to act as polarization sensitive antennas. Observing at longer wavelengths is OLIMPO (the Osservatorio nel Lontano Infrarosso e le Microonde su Pallone Orientabile) [71–74], a 2.6 m MKID telescope on a balloon that can reach altitudes up to 38 km. It is optimized to study galaxy clusters and observes in four bands around 150, 250, 350, and 460 GHz (2.0, 1.2, 0.86, and 0.65 mm). Its MKIDs are made from aluminum, it has between 19 and 41 pixels per band and it uses an MKID design similar to NIKA with a fractal inductor as sensitive antenna for all polarizations. OLIMPO has concluded its first observation campaign in 2018 after being launched from Longyearbyen, Svalbard and circling the North Pole. This flight was the first time MKIDs have been used for not-ground-based observations and was able to prove [72,74] that MKIDs are capable to withstand the demanding conditions on high altitude balloons. MKIDs from Al have also been part of the proposal for SKIP (Stratospheric Kinetic Inductance Polarimeter) [75], a balloon-born 0.5 m telescope to observe the cosmic microwave and infrared backgrounds. SKIP intended to have 2317 single-polarization MKIDs and be capable to observe in a band either around 150 GHz (2.0 mm) or between 260 and 350 GHZ (1.2–0.9 mm), depending on installed filters. Finally, MKIDs are also intended to be used with B-SIDE [76], a further proposed, 0.8 m aperture balloon-born instrument to observe polarized emission between 400 and 700 GHz (750 and 430 μm) with 980 to 1800 pixels, meant to improve foreground subtraction for CMB B-mode observations.

Ground-based observations of the Cosmic Microwave Background (CMB) have been a significant driving force behind the development of low temperature detectors for many years. Especially the observation of the so-called B-mode polarization of the CMB is attracting tremendous scientific interest as it could provide evidence for cosmic inflation shortly after the Big Bang. There are multiple CMB B-mode experiments ongoing and in late planning phases at the moment, and MKIDs are considered for several of them and already utilized in some. GroundBIRD [77–79] is searching for the CMB B-mode polarization with a 30 cm telescope that rotates rapidly (about 20 rounds per minute) on a horizontal platform to enhance noise suppression. It uses 161 MKID pixels in two bands centered around 145 GHz (2.1 mm, 138 pixels) and 220 GHz (1.4 mm, 23 pixels) and will start observing from Teide Observatory, Tenerife soon. A similar concept of a rapidly rotating cryogenic CMB B-mode telescope has been proposed in a feasibility study for an instrument to be located at Isi Station, Greenland [80], utilizing 2300 MKIDs in bands around 150, 210, and 267 GHz (2.0, 1.4 and 1.1 mm). ACTPol, the polarization sensitive experiment at the Atacama Cosmology Telescope is developing multi-chroic MKIDs for CMB polarization studies [81]. BICEP-Array [82,83], a planned next generation CMB experiment with an array of four telescopes at the geographic South Pole is considering the use of membrane suspended MKIDs [84] for their 220/270 GHz band (1.4/1.1 mm), as the desired 22,000 detectors would otherwise be challenging to multiplex. CMB-S4, a proposed next-generation ground-based CMB experiment which plans to combine 21 telescopes and 550,000 detectors, is also listing MKIDs as a detector candidate [85] in its current design state.

### 3.2. Observational Astronomy: Near-IR, Optical and UV

Starting at around 1.5 µm wavelength [86,87] the amount of Cooper pairs broken by even a single photon gets big enough to generate sufficiently large signals to allow MKIDs to detect and count single photons. This energy limit can depend, among others, on the detector's geometry, the thickness of the superconducting layer or the pixel size and work is going on to push it further into the mid-IR [64]. For photons at or above this photon counting limit, MKIDs offer four main advantages over competing detector technologies for observational astronomy:

- No dark counts: Every photon detected by an MKID produces its own, individual signal. The only event that would produce signal spikes in the absence of photons (a dark measurement) is the impact of two or more lower energy photons (from e.g., IR black body radiation) on the detector within a time frame below its time resolution of a few µs. As these are statistically very unlikely MKIDs have virtually no dark counts. The photon flux measured by MKIDs is therefore precise especially in low light scenarios as it is only limited by statistical fluctuations of the optical system's quantum efficiency.

- Energy resolution: The signal produced by a single photon absorbed by an MKID depends on the number of Cooper pairs broken by the photon's energy. The more Cooper pairs are broken the bigger the signal gets. A statistical effect called the Fano limit restricts the achievable precision, but the photon's energy can still be reconstructed from its signal height with up to medium resolution. This allows MKID arrays to measure every detected photon's energy and therefore to offer a low to medium resolution spectrum for every MKID pixel in the detector array, without requiring refractive optics.

- High time resolution: As every detected photon produces its own signal pulse, and as every MKID in a detector array is monitored around once every µs it is possible to determine every photon's arrival time with a precision between one and a few µs. (Design considerations of MKID detectors, like for example very narrow resonances, can result in detector reaction times that are slower than the monitoring frequency, but rarely more than a few µs.) This extreme time resolution allows MKIDs to be used to observe fast, time varying events.

- Scalability: Other low temperature detector technologies also offer photon counting without dark counts, energy resolution, or high time resolution. But modern astronomical instrumentation often requires pixel numbers in the millions or beyond. The unique advantage of MKIDs is their great multiplexability, and with it their capability to scale up to large detector arrays, even in the mega-pixel range. This allows

MKIDs to overcome the severe limitation in pixel numbers that other superconducting detectors suffer from and makes it possible to utilize the significant low temperature detector advantages in modern astronomical instruments.

Millisecond pulsars are one of the fastest phenomena observable in astronomy, and some do emit in the optical. The Crab Pulsar for example pulsates with a period of about 33 ms, mainly in radio but also at shorter wavelengths. These time scales are difficult to resolve with standard CCD or CMOS cameras but pose no challenge to an MKID instrument that is read out much faster. The ARray Camera for Optical to Near-IR Spectrophotometry, ARCONS, is an MKID instrument that was used to observe pulsars [88–90], among them the Crab Pulsar, and demonstrated in collaboration with the Green Bank Observatory a correlation between optical pulses and so called giant radio pulses [89]. ARCONS is a 2024 pixels camera with MKIDs made from sub-stochiometric TiN$_x$, it is sensitive between 350 nm and 1100 nm and capable to reach an energy resolution of up to 10% [91,92]. It was commissioned in 2012 and used with the Hale Telescope at Palomar Observatory and the Shane Telescope at Lick Observatory. ARCONS is one of the earliest MKID instruments used for astronomical observations and was designed to demonstrate what MKIDs can offer for astronomy. It has been used, for example, to study transiting binaries [93] and to demonstrate optical imaging spectroscopy with MKIDs [91]. ARCONS has now been replaced by a similar instrument with a much bigger pixel number, DARKNESS [2,94].

DARKNESS, the DARK-speckle Near-infrared Energy-resolving Superconducting Spectrophotometer [2,94] is a second generation MKID camera for 800–1400 nm photons, using a 10,000 pixels array of MKIDs made from PtSi. DARKNESS has a spectral resolution $R = \lambda/\Delta\lambda$ of 5–7 and is designed to be used either as a general MKID camera or as a dedicated instrument for the high contrast imaging of exoplanets. The direct observation of exoplanets in high contrast with a coronagraph to block out the host star's light is one of the most impressive applications for optical to near-IR MKIDs in astronomy at the moment. If a coronagraph is used to block the light from a host star to allow to observe its planets, turbulences in the air columns above the telescope cause perturbations, so called speckles. To reach maximum contrast between star and exoplanet and to be able to observe planets nearer to their hosts, these speckles have to be eliminated as best as possible by a feedback system to a deformable mirror. As speckles are chromatic and move fast the quick readout and intrinsic energy resolution of MKIDs result in a much better feedback loop compared to standard CCD cameras. Besides, the fact that MKIDs have no dark counts allows additional statistical post processing procedures that in combination with their time and energy resolution advantages should allow MKIDs to increase achievable contrast ratios with coronagraphs by up to two orders of magnitude [2]. As this increase in contrast could be sufficient to enable the upcoming 30 m class of optical telescopes to observe rocky planets in the habitable zone of nearby stars the interest in optimizing coronagraphic contrast ratio is immense. DARKNESS is optimized to work behind the PALM-3000 extreme adaptive optics system and the Stellar Double Coronagraph at the Hale Telescope at Palomar Observatory and has been used as a technology testbed for high contrast imaging with MKIDs. It is planned [95] to also use DARKNESS with the MagAO-X system on the 6.5 m Magellan Clay telescope in Las Campanas, Chile.

Apart from DARKNESS there are currently two further MKID cameras being developed for coronagraphy. The Picture-C MKID camera [96] is similar to DARKNESS with a 10.000 pixels MKID array covering 540–660 nm. It is intended to be used with the second flight of the Picture-C (Planetary Imaging Concept Testbed Using a Recoverable Experiment—Coronagraph) project [97], a high altitude balloon with a coronagraph behind a 0.6 m telescope. Picture-C is optimized to image debris disks, exozodiacal light and young hot-Jupiters around nearby stars and is used as a testbed for future Earth-like exoplanet imagers. The currently largest MKID array in use is MEC, the MKID Exoplanet Camera [3,98]. MEC is an integral field spectrograph with a detector array of 20,440 MKIDs made from PtSi, is sensitive between 800 nm and 1400 nm and reaches a spectral resolution of 5–7. MEC has been integrated with the Subaru Coronagraphic Extreme Adaptive Optics (SCExAO) instrument [99] at the Subaru 8 m telescope on Mauna Kea, Hawaii in 2019 and is expected to significantly improve star to planet contrast ratios. MEC is planned to be made available for open use observations in the near future.

The inherent energy resolution MKIDs offer can obviously be useful for integral field spectrographs (IFS). An IFS is basically a camera where every picture element provides its own spectrum. With current Si-based detectors this can be achieved by using several filters or stretching the light of every object of interest into a line spectrum across the detector with diffractive optics. The filter approach has a low energy resolution equal to the number of filters and is very slow. Projecting multiple line spectra next to each other onto a CCD array is used very successfully but requires complex optics and has its limits when too many objects are present as lines start to overlap or in the observation of extended sources. Even though optical to near-IR MKIDs can so far only reach a low to medium energy resolution they still allow to significantly enhance the performance of IFSs in dedicated situations. Giga-Z [100–102] is a proposed MKID wide-field multi-object spectrograph for the 350–1350 nm range and promises to significantly improve LSST follow-up observations if deployed behind a 4 m class telescope and if an MKID array of 100,000 pixels can be used. (LSST, the Large Synoptic Survey Telescope, recently renamed the Vera C. Rubin Observatory, is a 8.4 m telescope under construction on Cerro Pachón, Chile. It will monitor the night sky constantly and is expected to detect an enormous amount of transient astronomical events like e.g., supernovae.) A more recently proposed IFS is KRAKENS, the Keck Radiometer Array using KID ENergy Sensors [102–104]. Krakens is planned to be a facility class IFS for the Keck-1 telescope on Mauna Kea with a 30,600 MKID pixels array, a 42.5″ x 45″ field of view and a sensitivity range between 380 nm and 1350 nm. In combination with a desired improvement of MKID spectral resolution to R = 20 this would enable a broad range of novel observational opportunities, and a proposed update to 57,600 pixels and a 1″ x 1″ field of view would further enhance these. It is clear that an MKID IFS needs to have a large number of pixels to be able to compete with existing technologies and that it would profit significantly from further progress in MKID single pixel energy resolution.

A further possibility to utilize the capability of MKIDs to resolve the energy of individual detected photon is KIDSpec [102,105,106], an MKID instrument currently in development at Durham University. KIDSpec, the Kinetic Inductance Detector Spectrograph is a highly efficient, medium spectral resolution, wide band-pass astronomical spectrograph. The idea behind KIDSpec is to use a linear MKID array as photo sensitive element and at the same time as order resolver for a low line density echelle grating. This way the MKID detectors replace the cross-disperser in a classical echelle spectrograph, greatly simplifying the optical layout, reducing optical losses from a second grating and allowing longer slits. KIDSpec is expected to achieve a spectral resolution R = 4000–10,000 in a 400–1500 nm band and to have better sensitivity for low intensity lines as MKIDs have no dark counts and allow better sky background and cosmic ray subtraction.

Further proposals to utilize photon counting MKIDs in astronomical instrumentation include a general push towards better miniaturized instruments [107] and a Gaia type satellite to map the Milky Way in the near-IR [108] (Gaia is a space observatory by the European Space Agency. Its mission is to measure the precise position, luminosity and motion of over one billion stars in the optical wavelength range). This would give better results near the galactic center and some spiral arms compared to GAIA as IR is blocked much less by dust compared to optical light. MKIDs are one option for this kind of observation and are attractive as their fast readout allows for easier and faster all-sky mapping. There are also development efforts going on to adapt the MKID detection principle in order to be able to detect soft and hard X-Rays and medium to high energy particles [109–112]. As at higher energies, absorbing the particle or photon to be registered in the detector starts to get challenging this requires use of so called Thermal Kinetic Inductance Detectors (TKIDs). A TKID operates as micro-calorimeter as it has a dedicated particle absorber and the temperature rise of that absorber results in Cooper pair breaking in the TKID's inductor. As TKIDs have the same multiplexability advantage as MKIDs they promise scalability up to energy resolving, mega-pixel arrays. They are still in an early prototype state but the possibility to achieve high pixel numbers makes them a promising future detector candidate for applications both in astronomy and for synchrotron light sources.

## 4. MKIDs in Particle Physics

The above might lead the reader to believe that MKIDs are only useful for astronomical applications. Instead, Microwave Kinetic Inductance Detectors are widely employed in the context of particle and astro-particle physics. Some research groups are studying the feasibility of using MKIDs for rare events experiments such as the experimental evaluation of the neutrino mass and the detection of Dark Matter particles such as WIMPs (Weakly Interacting Massive Particles). In both cases, MKIDs based instruments have been proposed both as means of direct detection and to further reduce the background signal of current experiments.

### 4.1. MKIDs for Neutrino Physics Experiments

Particle physics experiments require an ever-improving level of technology to detect exceedingly rare and ever more difficult to find events. In particular neutrino physics is now a trending topic among many research groups. The possibility of using cryogenic detectors for neutrino physics experiments has been proposed and investigated since the 1980s [113]. The first project that investigated the possibility of deploying MKIDs for neutrino experiments was MARE [114], which unfortunately was cancelled ahead of time after a disappointing phase, MARE-1, which used TES detectors and Si thermistors [115]. Since then, MKIDs have significantly improved their performances and are under test for experiments such as CALDER [116], CUORE [117] with its planned upgrade CUPID [118], and HOLMES [119]. The proposed working scheme employs MKIDs as photodetectors in addition to neutron transmutation doped (NTD) Ge thermistors in order to suppress the background signal produced by spurious excitations and possible $\alpha$-decay of the radioactive medium.

#### 4.1.1. CUORE

CUORE (the Cryogenic Underground Laboratory for Rare Events) is an underground experiment based in LNGS (Laboratori Nazionali del Gran Sasso), Italy whose main task is to detect neutrino-less double beta decay events in refrigerated tellurium dioxide (TeO$_2$) crystals [120]. The experimental setup consists of 988 TeO$_2$ crystals for a total mass of 741 kg (206 kg of $^{130}$Te). The detectors deployed on this experiment are NTD Germanium thermistors [121] whose task is to detect temperature changes induced by the radioactive decay of the $^{130}$Te nuclei, which is a double-beta-decaying isotope of tellurium. For CUORE, just as for other rare event experiments, an optimal background rejection is crucial. It is expected that CUORE will have a background of $10^{-2}$ counts/kg/keV/year in the energy range of the neutrino-less double beta decay; this is expected to be approximately 200 counts over the 5 years the experiment will run for [122].

The successor of CUORE is currently being planned and will go under the name of CUPID (CUORE Upgrade with Particle IDentification). In addition to CUORE's current structure, CUPID intends to switch to Li$_2$MoO$_4$ crystals and will have particle identification capabilities which arises from the coupling of photodetectors to CUORE's bolometers. These detectors will allow to distinguish events due to $\alpha$-particles from the natural background (mainly decays in Copper) and double-beta decays as in Li$_2$MoO$_4$ alpha- and beta-particle events differ in light yield and time development of their scintillation light [123–126]. Although the photodetectors that will be deployed in CUPID are currently planned to be based on NTD Ge thermistors, multiple further options are being considered for possible future upgrades, one of which involves MKIDs as photo-detectors.

#### 4.1.2. CALDER

CALDER (the Cryogenic wide-Area Light Detectors with Excellent Resolution) is an experiment based in Rome and LNGS and its main goal is to develop sensitive MKIDs in order to identify rare events, in particular the group is working on answering two different fundamental questions [127]

(1) Is the neutrino a standard particle or is it its own anti-particle and therefore a Majorana particle? This information can be obtained by the detection of the particular rare neutrino-less double beta decay.

(2) What is the nature of the dark matter that fills our universe?

CALDER uses CUORE's former experimental setup with $TeO_2$ to develop large area phonon mediated MKIDs whose main task is the detection of ultraviolet and visible Cherenkov photons emitted in the $TeO_2$ crystals. The detection of such photons would allow to significantly improve background rejection: $\alpha$-particles from natural background radiation do not produce Cherenkov radiation in $TeO_2$ crystals, while electrons produced by a double-beta decay of Te would. The correct detection of said Cherenkov light could therefore be beneficial to distinguish the nature of these two different events that otherwise produce very similar signals.

CALDER is investigating the feasibility of fabricating phonon mediated MKIDs out of Al/Ti/Al tri-layers whose critical temperature can be tuned accurately by varying the thickness of the three superconducting layers. These devices produced an improvement in the RMS baseline resolution from 105 eV down to 26 eV, getting close to the target resolution of 20 eV [128]. Particularly interesting results have been shown by Casali et. al. [126] who utilized MKIDs to simultaneously detect photons and phonons in $Li_2MoO_4$ crystals and demonstrated that their MKIDs can achieve very high detection speeds and immense sensitive areas. This combination is very promising for further background reduction as it allows to reject pile-up events of 2-neutrino double beta decays, a potentially significant background in $^{100}$Mo based experiments [126]. Currently, the CALDER collaboration is working on further improving these promising results and increasing the sensitive area of the detector from 2 cm x 2 cm to 5 cm x 5 cm while also working on developing further detectors such as BULLKID [129], which we will describe later. It is also important to note that other materials such as $TiN_x$ and Ti/TiN multilayers are being taken into consideration for the sake of the optimization of the CALDER detectors [130,131].

#### 4.1.3. HOLMES

HOLMES is an experiment aimed at the direct measurement of the electron neutrino mass produced by electron capture decay of radioactive holmium nuclei $^{163}$Ho. The experiment is set up to measure the energy freed by the decay of $^{163}$Ho, detecting all the energy except for the mass of the neutrino. This simple calorimetric measurement is extremely powerful as it is completely independent on the model used to describe the physics of the neutrino particle. HOLMES is currently deploying Transition Edge Sensors provided by NIST, which are frequency domain multiplexed by coupling each TES to a SQUID (Superconducting QUantum Interference Device) and a readout MKID [132] to create an array of resonant circuits with unique resonant frequencies. Unfortunately, TESs are not ideal detectors for future upgrades of the experiment; even with frequency domain multiplexing the scalability of the array size beyond 1 kilo-pixels is difficult and the fabrication of such arrays could be a major challenge. In order to prepare for future upgrades to HOLMES MKID arrays are therefore under development. Their scalability to Mega-pixel numbers is significantly easier and their fabrication is also simpler as MKIDs are built on few superconducting layers. The cryogenic setup and read-out electronics of MKIDs being significantly simpler than that of TESs is another argument in favor of the technology switch. A Mega-pixel MKID array can yield sub-eV sensitivity for this experiment in the energy range of interest [133].

The MKIDs being developed for HOLMES are meant to be X-Ray detectors; typically, these detectors consist of an absorber coupled to a micro-resonator. The absorber, tantalum (Ta) in this case, having a high stopping power, absorbs X-Rays and produces quasi-particles which then propagate to the micro-resonator where they are efficiently trapped, and a detectable signal is produced. MKIDs can also be operated as thermometers (TKIDs), an increase in temperature could produce quasi-particle excitations the same way a photon would. This second approach exploits the thermal absorption of X-Rays in the absorber rather that the ionization of the same. Both detecting mechanisms are currently under investigation for the sake of the HOLMES experiment [133]. For this purpose, MKIDs made of Ti/TiN multilayers are being fabricated. Through the stacking of multiple layers of Ti and TiN the critical temperature of the superconductor can be tuned through the proximity effect with great accuracy from 70 mK to 4.6 K [133]. In the process of the optimization of TKID detectors, Ti/TiN multilayers micro-resonators have been coupled to a gold (Au) absorber deposited on a suspended silicon nitride ($SiN_x$) membrane. The performance of these detectors is still under investigation [134].

### 4.2. MKIDs for Dark Matter Experiments

WIMPs are a possible dark matter candidate, but unfortunately their interaction rate, usually intended as interaction with some nuclei, is extremely low, <0.1 event/kg/day and the expected energy deposited per interaction is in the order of 10 keV. State of the art experiments which are trying to detect WIMPs exploit phonon and photon detectors to measure the ionization of the interacting medium as well as the phonon excitation in the medium to try and distinguish electron recoils from nuclear recoils, which might be due to WIMPs. Further improvements are being pursued in order to increase the sensitivity of the detectors to WIMPs and to further reduce unwanted background signals. Improvements to the state-of-the-art detectors are intended to go in the direction of increasing the sensitive area, increased energy collection and fabrication reliability. In this framework MKIDs can be a means to achieve such desired improvements given their sensitivity, their multiplexability and their insensitivity to inhomogeneities in critical temperature which, in return, coincides with a much easier to fabricate detector.

To this day, only few applications for MKIDs to be deployed as dark matter detectors have been proposed. S. Golwala pioneered the idea [135] in 2008, and prototype detectors have been developed by Caltech and JPL [136,137] based on multiple MKIDs spread on a massive absorber. Phonons caused by particle absorption would trigger most of them and time delays and signal intensities allow to localize the interaction location. In 2020 the idea was followed up by the CALDER group in LNGS which published their white paper on BULLKIDs, a R&D project to exploit MKIDs for the detection of rare, low-energy processes [129]. Golwala et al. imagined the interacting medium to be the silicon substrate on which the MKIDs are fabricated. The interaction would excite phonons which are then trapped in the superconducting films exciting quasi-particles. The resulting quasi-particles are eventually trapped in the MKIDs provided that they are fabricated with a superconductor whose band gap is smaller than that of the absorbing material. In particular, the proposed instrument would be produced with a Ta phonon absorber and Al MKIDs. Said instrument would be expected to have an energy resolution of 58 eV for a 20 keV event in the detector.

The BULLKID devices proposed by Colantoni et al. in [129] instead use 108 individual 5 × 5 × 5 mm³ Si voxels weighing 0.29 g each as the interacting medium. These voxels are fabricated on the backside of the Si wafer where the MKIDs will then be structured. The primary goal of BULLKID is the high-precision measurement of coherent elastic scattering of neutrinos off nuclei and could also be employed as a dark matter detector [129]. Dark matter WIMPs are intended to interact with the silicon substrate producing phonons that the MKIDs on the other side of the wafer is expected to detect. In particular, the fact that the voxels are separated one from another allows for a rough, but immediate, determination of the position where the interaction might have occurred. The superconducting material intended to be used for these BULLKIDs detectors is a multilayer stacking of Al/Ti/Al whose critical temperature can be tuned to preference. A $T_c$ of 805 mK is expected for tri-layers whose respective thickness are 13, 33, and 30 nm. An optimization of these devices is projected to allow for an energy resolution as low as 20 eV.

In addition to the more canonical approaches just described, two further applications are being investigated. The detection mechanism remains mostly unvaried: weakly interacting massive particles (WIMPs) need to recoil off other particles. This interaction, in return, produces excited particles whose energy can be measured. The paper by Hochberg [138] proposes a calorimetric approach in which the interacting medium is a superconductor in which the dark matter particle interacts with a free electron at the Fermi energy, thus producing quasi-particles which can then be detected with the help of MKIDs. These proposed detectors are uniquely suited to detect dark matter particles whose mass is sub-MeV. Oppositely, Derenzo et al. [139] investigates the possibility to use MKIDs coupled to scintillating materials such as sodium iodide (NaI), cesium iodide (CsI), and gallium arsenide (GaAs). Detecting scintillating photons rather than phonons can allow for measurements with significantly lower backgrounds, as there are no Compton scattering events to rule out. Clearly, this detecting mechanism is not background-free since effects such as "afterglow", phosphorescence induced from previous interactions, can create spurious photons in the scintillating material. But phosphorescent photons typically have a lower energy than those produced by direct scintillations. MKIDs are especially suitable for this application as they are inherently energy resolving detectors and are, for most practical purposes, effectively dark count free.

Along with WIMPs, axions and dark photons are plausible dark matter candidates whose mass range extends down to fractions of one eV. An MKID based experiment has been proposed to detect axions whose masses are in the range of 0.25–2.5 eV [140,141]. It proposes a photon-mediated detection mechanism for such elusive particles. Through the interaction of axions with optical structures, such as stacks of dielectrics, whose structure's period is of the order of the dark matter's Compton wavelength, the energy of the dark matter particles can be converted into relativistic photons that can be effectively detected [142]. Said photons are emitted in a direction perpendicular to the surface of the stack with a very small ($10^{-3}$ degrees) aperture angle [142]. The focusing of the thus-produced photons on a detector array can be easily performed with an appropriate lens. MKIDs stand out among the possible detectors to be used for such applications thanks to their low-noise levels, the virtual absence of dark counts and high integrability [140].

## 5. MKIDs Applications in Material Science with Synchrotrons

Another field where kinetic inductance detectors are being used is in the use of synchrotron radiation for material science. Synchrotron radiation is emitted by electrons radially accelerated to speeds close to the speed of light in the vacuum of a particle accelerator. In the reference frame of the electron, this emission is at radio frequencies, while in the reference frame of the detectors, it is shifted into the X-ray range. A periodic series of magnets, called undulators, cause transverse undulations in the electron. The wavelength of the emitted radiation is a function of the period of these undulators. Synchrotrons have increased the brightness of X-ray sources by 26 orders of magnitude since the 1970s. Using MKIDs as detector element in synchrotron beam lines allows to exploit their major advantages of energy resolution and their ability to be multiplexed to large arrays, and only having to deal with the disadvantage of requiring advanced cryogenics to cool them down to mK temperatures [143].

T. Cecil et al. [110] describe developing MKIDs for synchrotron X-ray spectroscopy, citing their ability to scale up to large arrays. They list particular uses of this X-ray spectroscopy as synchrotron experiments, laboratory tools and astronomical satellites. They proposed WSi$_x$ for these detectors due to it having a short attenuation length for soft X-rays and controllable $T_c$ in the range of 1 to 3 K, and it being able to be sputtered from separate W and Si targets with good control. Exact values for $T_c$ can be tuned by changing the W to Si ratio. Initial results show quality factors greater than $10^6$ [144].

They later instead suggest [112,145] using TKIDs in a thermal quasi-equilibrium state to detect X-ray photons. They use TKIDs as they do not require as strict limitations on the absorber design, as regular MKIDs for X-rays would. In order to assess the ability of TKIDs to be used for X-ray spectroscopy they have simulated their operation with a range of parameters, relevant to measuring X-ray photons, and examined the detector design parameter space. The response of the simulated TKID was then tested with several noise sources, and these simulated results were then used to design new devices. Based on results from these simulations, optimized TKIDs were designed for 6 keV X-rays. This optimized TKID is proposed to have a $T_c$ of 1 K, heat capacity of 1 pJ/K, and a resonator bandwidth of 6 kHz FWHM. They suggest that further improvements could be made with the use of a parametric amplifier, as the HEMT amplifier noise was the limiting factor.

Ulbricht et al. [109] have also developed TKIDs for X-ray imaging spectroscopy, stating potential applications of both X-ray astronomy as well as X-ray spectroscopy at synchrotron radiation sources. While diffractometers can already realize energy resolution of ~0.5 eV for 5.9 keV [146] photons, they have to be scanned across their energy range, which TKIDs do not. While TESs are also being used for spatially resolving X-rays, showing energy resolutions of 2.4 eV at 5.9 keV, and pixel counts of over a thousand, they run into difficulties when scaling up to larger arrays. In [109] TKIDs have been fabricated from sub-stoichiometric TiN$_x$ films, a Nb feedline and ground plane, and a 500 nm thick Ta absorber, isolated from the inductor by a free standing Si$_3$N$_4$ membrane. With a critical temperature of 1.6 K, and an operating temperature of 75 mK to 200 mK, temperature rises of 200 mK to 300 mK were measured from 6 keV photons. Exciting the fabricated TKID with an Fe$^{55}$ source, 4970 photons were plotted as a histogram. From this, an energy resolution of 75 eV was calculated at 5.9 keV. Ulbricht et al. [109] conclude that further optimization should be able to improve the TKID energy resolution to that achieved by TES calorimeters.

Faverzani et al. [134,147] are developing TKIDs for X-ray spectroscopy in the keV range with the potential future application of measuring the mass of neutrinos. These TKIDs are made of Ti/TiN multilayers, with an absorber suspended on an Si$_3$N$_4$ membrane. Three different films were produced, with Ti/TiN thicknesses of 10/7 nm (total thickness of 100 nm), 10/10 nm (total thickness of 102 nm), and 10/12 nm (total thickness of 110 nm), giving critical temperatures of 0.6 K, 0.8 K, and 1.2 K respectively and kinetic inductances of 30 pH/□, 20 pH/□, and 12 pH/□. Each chip produced contained four resonators, each comprising of two interdigitated capacitors and an inducting line which acts as the absorber. These resonators have been tested successfully, with quality factor $Q_c$ = 17,000 and $Q_i$ = 82,000, at a resonant frequency of 5.598 GHz. Signal pixel optimization is being carried out, after which the best performing resonator design will be implemented in a larger array.

## 6. MKIDs for Security Applications

A further area for which the use of MKIDs is being investigated is security applications. Currently, there are groups based in Cardiff, Glasgow, Rome, Helsinki, and Groningen that are developing passive GHz and THz imaging cameras based on MKIDs, for security applications such as airport screening, the detection of hidden objects such as landmines, and the screening of trucks and vans. MKIDs are a promising technology for these applications because they allow large arrays of detectors due to their multiplexabiliy, with time responses of ~ $10^{-4}$ s, while detecting frequencies from about 300 to 1000 GHz (1.0-0.3 mm), at which certain materials which are opaque at optical wavelengths become transparent. While active THz imaging is already used in airports around the world for security screening, MKIDs would, due to their improved sensitivity, allow for passive THz imaging with fast time responses, allowing for subjects to be scanned while simply walking past the camera, greatly reducing the time needed for each traveler to be screened [148–152].

Doyle et al. in Cardiff University have developed a passive 350 GHz (860 μm) field-scanning LEKID camera [148], operating at a "quasi-video" frame rate of 2 Hz, and noise equivalent temperature of ~0.1 K per frame based on a linear array of 152 LEKIDs. The array is fabricated from an aluminum film deposited onto a high purity silicon wafer. This camera successfully demonstrated sufficient spatial resolution, sensitivity and scanning speed to detect concealed objects in a person's clothing or luggage. The array is housed in a cryostat with a 25 cm diameter window, into which incoming light is coupled using lenses and a flat beam forming mirror. At any one moment, a thin, horizontal strip of the object is scanned with the mirror, while oscillating the mirror allows the full object to be covered. This array achieved a pixel yield of 85%, with the unusable pixels being attributed to non-uniformity in the aluminum film thickness, causing overlapping resonators. This LEKID array was used to detect objects hidden inside a subject's coat, like a wallet, an air pistol, and some coins.

Hassel et al. [151] in Helsinki have developed a dual-band imaging system, CONSORTIS, based on 8208 kinetic inductance bolometers (KIBs) in an array. This is also being developed to be used as a passive, "walk by" scanner for security screening purposes. KIBs are equivalent to TKIDs thermal detectors, consisting of an absorber and inductive thermometer on a thermally confined membrane. This thermometer is based on the temperature dependence of the kinetic inductance effect, and is coupled to an external capacitor, creating a resonance. The KIBs are then read out in the same manner as TKIDs. One third of the 8208 pixels are designed to operate at a frequency band centered at 250 GHz (1.2 mm), while the remaining two thirds of the pixels are sensitive around 500 GHz (600 μm). This system has demonstrated detection of an aluminum bar as well as plastic and paper objects underneath a person's jacket. They report that they demonstrated the basic operation of this system at detecting concealed objects, and that work is ongoing to scale up this technology to allow all the detectors in the array to be read out simultaneously. It is reported that an issue with proper thermalization of the full focal plane has been resolved, and that work is ongoing to fully parallelize the readout.

Messina et al. [150] in Rome have designed and fabricated high $T_c$ MKIDs, based on YBa$_2$Cu$_3$O$_7$ (YBCO), reaching a critical temperature of up to 89 K, for use in security applications. They aim to use high $T_c$ MKIDs to allow the use of cheaper, more reliable cryogenic systems, needing to only reach temperatures of 20 to 40 K, as opposed to typical mK temperatures. They propose that an MKID made of high critical temperature superconductor (HCTS) would give the high sensitivity of MKIDs, with the cost and reliability benefits of Stirling coolers. Simulations have been carried out on a proposed design for these HCTS-MKIDs for THz imaging. An electron beam lithography process has been developed for the fabrication of these devices.

Morozov et al. [149] in Glasgow have investigated the use of atomic layer deposition (ALD) to deposit titanium nitride (TiN) thin films, for use in MKIDs for passive THz imaging. TiN thin films were the material of choice due to their greater than 300mK critical temperatures, allowing the use of compact cryogenic systems, reducing the overall complexity of the system. They report that for modern security applications, frame rates of 25 Hz and noise effective temperature differences (NEDT) of ≤0.1 K are desired, and that for a system similar to that reported by Rowe et al. [148] a time constant τ of ≤100 μs and a noise effective power, *NEP*, of the order of 100s of pW is needed. A test array consisting of 12 pixels, made up of an inductive meander, in series with an interdigitated capacitor, coupled to a 50 Ω coplanar waveguide, was designed and fabricated. They describe results from this test array of *NEP* ≈ 2.3 × 10$^{-15}$ W/√Hz, and τ ≈ 31 μs. This time constant meets the stated requirement, but the *NEP* is higher than the photon noise of the source, defined by the 250 K blackbody temperature of the test source. They suggest that optimization of the pixel design, and reduction in the readout noise is needed to improve this figure.

De Jonge et al. [152] in Groningen are also developing a THz imager for passive, "stand off" imaging for security and biomedical applications. They describe an instrument based on an MKID array, cryostat, warm and cold optics, and mechanical scanner. As an initial demonstration of the use of MKIDs for passive screening, they developed a 4 × 4 pixel array for the 350 GHz frequency band (860 μm). An external scanner was used to create full images of 70 × 70 pixels with spatial resolution of 3.2mm of a 220 × 220mm area. This scanning system was then used to take the first THz MKID images of a biological subject, namely, the hand of the author.

## 7. Further MKID Applications

MKIDs are still a relatively new technology whose capabilities have not yet been exploited to the fullest, partly because the optimization of these innovative detectors is still ongoing. As described through the length of this review paper, Microwave Kinetic Inductance Detectors are being deployed intensively in astronomical applications. Further scientific applications already discussed include particle and astro-particle experiments, and material science, in particular as synchrotron radiation detectors. MKIDs also showed promising results for practical applications such as security scanners for passengers at airports and freight at harbors. In this last chapter we want to present further applications of Microwave Kinetic Inductance Detectors which have been proposed or are at the first stages of investigation but did not fit well within the areas already discussed.

A preliminary study investigating the possibility of using polarization-sensitive MKIDs for plasma diagnostics has been published by F. Mazzocchi et al. [153]. Polarimetric measurements allow for the determination of various plasma parameters. Usually, polarimetric systems are based on gas lasers emitting in the THz range which pose many issues with reliability, safety, and stability. A possible alternative approach involves a Quantum Cascade Laser, but due to its low power this new source comes with a need for an extremely sensitive detector, such as MKIDs [153].

Another promising application for Microwave Kinetic Inductance Detectors is the development of a Quantum-Limited Atomic Force Microscopy (QAFM) [154]. Such an ambitious project requires the development of electro-mechanical sensors which are capable of performing measurements at the quantum limit. The designed sensor would exploit the mechanical strain in the superconductor to create quasi-particles and vary the kinetic inductance of the resonator. By monitoring the superconducting resonator, it is then possible to identify the resonant frequency of the cantilever and therefore produce an image of the scanned surface. Further studies on the behavior of the resonator under strain are ongoing; in order to produce vibrations a piezoelectric shaker is being included in the experimental setup [154].

**Author Contributions:** Chapters 4 and 7, sketches of Figure 1: M.D.L.; chapters 5 and 6: E.B.; conceptualization, supervision, chapters 1–3, review and editing: G.U. All authors have read and agreed to the published version of the manuscript.

**Funding:** This research was funded by Science Foundation Ireland under Grant number 15/IA/2880.

**Acknowledgments:** The authors would like to thank Tom Ray, Dublin Institute of Advanced Studies, for making this work possible and many helpful discussions.